\documentclass[sigconf]{acmart}

\AtBeginDocument{%
  }

\usepackage{url}

\usepackage{amsmath,amssymb,amsfonts,amsthm}
\usepackage{algorithmic}
\usepackage{graphicx}
\usepackage{textcomp}
\usepackage{listings}
\usepackage{multirow}
\usepackage{float}                 
\usepackage{subfig} 
\usepackage{tabularx}
\usepackage{braket}
\usepackage{hyperref}

\setcopyright{acmlicensed}
\copyrightyear{2018}
\acmYear{2018}
\acmDOI{XXXXXXX.XXXXXXX}

\begin{document}

\title{QSolver: A Quantum Constraint Solver}

\author{Shangzhou Xia}
\email{xia.shangzhou.218@s.kyushu-u.ac.jp}
\orcid{0009-0006-2775-9633}
\affiliation{%
  \institution{Kyushu University}
  \city{Fukuoka}
  \country{Japan}
}

\author{Haitao Fu}
\email{fu.haitao.968@s.kyushu-u.ac.jp}
\affiliation{%
  \institution{Kyushu University}
  \city{Fukuoka}
  \country{Japan}
}

\author{Jianjun Zhao}
\email{zhao@ait.kyushu-u.ac.jp}
\affiliation{%
  \institution{Kyushu University}
  \city{Fukuoka}
  \country{Japan}
}
\authornote{Corresponding author}

\renewcommand{\shortauthors}{Shangzhou Xia, Haitao Fu, Jianjun Zhao}

\begin{abstract}
\textcolor{black}{With the growing interest in quantum programs, ensuring their correctness is a fundamental challenge. Although constraint-solving techniques can overcome some limitations of traditional testing and verification, they have not yet been sufficiently explored in the context of quantum programs. To address this gap, we present QSolver, the first quantum constraint solver. QSolver provides a structured framework for handling five types of quantum constraints and incorporates an automated assertion generation module to verify quantum states. QSolver transforms quantum programs and multi-moment constraints into symbolic representations, and utilizes an SMT solver to obtain quantum states that satisfy these constraints. To validate the correctness of the generated input states, QSolver automatically generates assertion programs corresponding to each constraint. Experimental results show that QSolver efficiently processes commonly used quantum gates and demonstrates good scalability across quantum programs of different sizes.}
\end{abstract}




\maketitle

\section{Introduction}
Quantum programs are increasingly used in various fields due to their unique computational advantages. Unlike classical programs, quantum programs leverage quantum superposition and entanglement to perform computations~\cite{nielsen2010quantum}. 
\textcolor{black}{However, due to the intrinsic characteristics of quantum mechanics, the outputs of quantum programs are inherently probabilistic, and the intermediate execution states remain unobservable. Although measurement enables the acquisition of the probability distribution of program outcomes, it inevitably collapses the current quantum state. These pose significant challenges in debugging and ensuring the correctness of quantum programs~\cite{Qroadmap, testing-debugging,Qformal-method}.}


Recognizing these challenges, researchers have proposed test-based approaches~\cite{ontesting,quratest,blackbox,long2024testing,10.1145/3728926,property-based, mutation, coverage,jin2025,projection-based,SSBSE} and formal verification methods~\cite{10.1007/978-3-031-27481-7_12,10.1145/3656419, silq_smt, formal-error,logic-formal,hoare-formal,QWIRE} to improve the correctness of the quantum program. Test-based approaches rely on diverse test cases but often lack sufficient exploration, whereas formal methods, although rigorous, are typically inflexible and 
\textcolor{black}{their emphasis on verifying predetermined inputs limits the applicability to integration testing.}
Furthermore, to verify constraints at different moments within a quantum program, mid-circuit measurements are required. However, such measurements collapse the quantum state, making it impossible to execute subsequent quantum operations. This limitation hinders existing testing approaches from simultaneously enforcing constraints across different moments of quantum operations.




\textcolor{black}{Constraint solving approaches~\cite{dreal,z3,cvc5}, owing to their exhaustive exploration capabilities and flexible modular analysis, have been widely applied in classical programs. However, their application in quantum programs has not yet been sufficiently explored. To bridge this gap,}
we present QSolver, the first quantum constraint solver designed to facilitate the testing and verification of quantum programs. Implemented in Python, QSolver utilizes the symbolic representation of quantum operations in SymQV~\cite{10.1007/978-3-031-27481-7_12}. QSolver provides five types of quantum constraints specifically for measurement results and generates automated assertion programs to verify solutions. 
\textcolor{black}{To ensure that the output satisfies constraints across different moments, QSolver encodes all quantum operations and constraints from various execution moments into a unified SMT file for joint solving.}
\textcolor{black}{This modular generation approach can handle constraints at any moment within a quantum program.}


\textcolor{black}{We evaluate QSolver through two key experiments: (1) its capability to handle diverse types of quantum gates, and (2) its ability to process quantum programs of different sizes. Experimental results demonstrate that QSolver effectively manages complex quantum constraints and handles a wide range of quantum gate operations.}


\section{Preliminaries}
\label{sec:preliminaries}

\begin{figure}[ht]
\centerline{\includegraphics[width=0.8\linewidth]{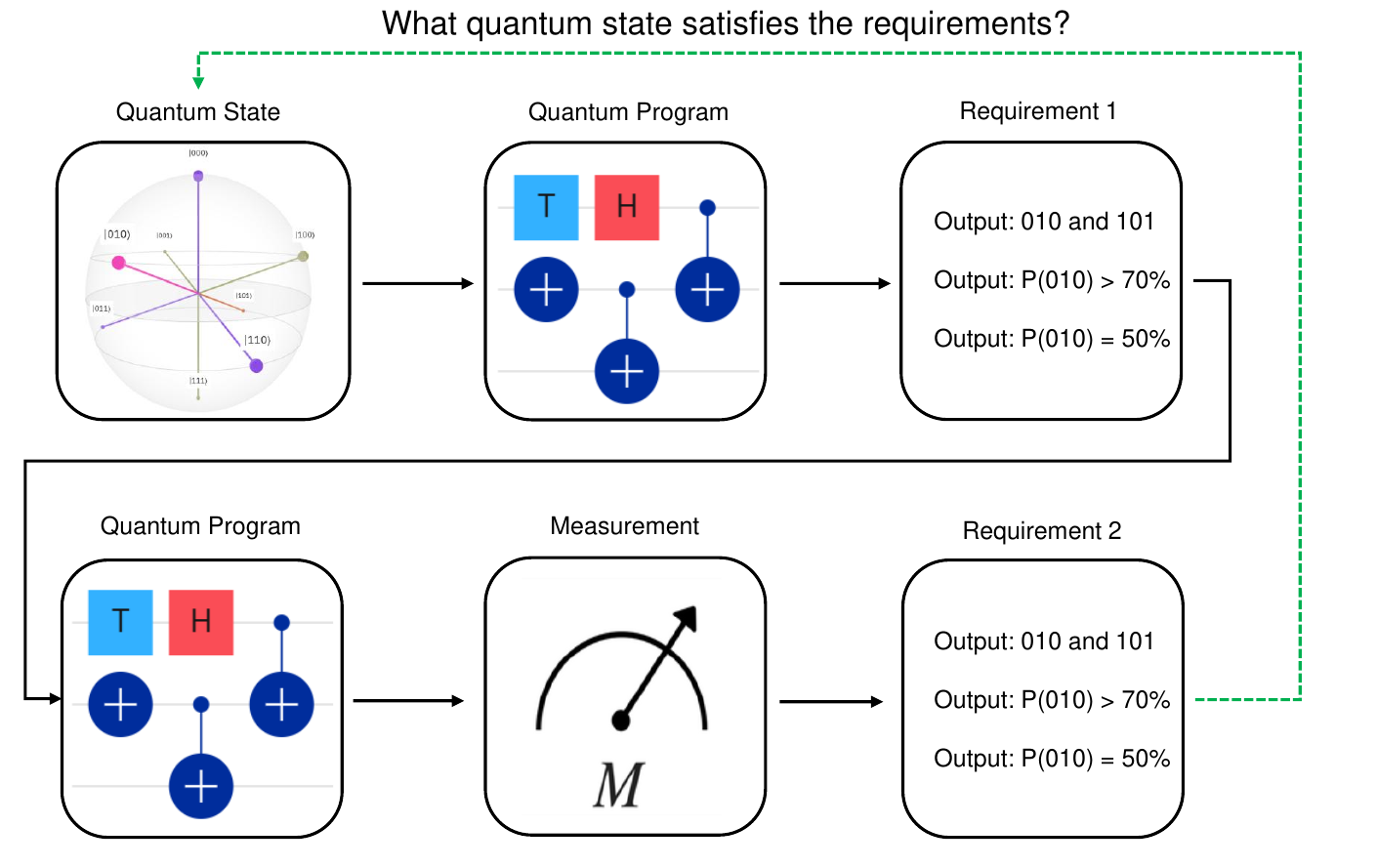}}
\vspace{-2.5mm}
\caption{\textcolor{black}{A motivation example: The execution flow of a quantum program begins with an input quantum state. The state undergoes a series of quantum gate operations, and the intermediate state is subjected to requirement 1. Then, the program proceeds with further quantum operations, and finally, measurement is performed to obtain the output result along with its corresponding probability requirement 2.}
}
\label{fig:motivation}
\vspace{-4mm}
\end{figure}

\textcolor{black}{This section introduces some background, including quantum states, quantum gates, and a motivation example, as illustrated in \autoref{fig:motivation}.}

\subsection{\textcolor{black}{Quantum State}}
A quantum state represents the physical state of a quantum system, typically described using qubits. Unlike classical bits, which take discrete values (0 or 1), a qubit can exist in a superposition of states, mathematically expressed as a unit vector in a complex Hilbert space. For a single qubit, the state is represented in Dirac notation as:
$\ket{\phi} = \alpha_0 \ket{0} + \alpha_1 \ket{1}$, where \(\alpha_0, \alpha_1 \in \mathbb{C}\) are probability amplitudes satisfying the normalization condition \(|\alpha_0|^2 + |\alpha_1|^2 = 1\). For an \(n\)-qubit system, the composite state is described using tensor products:$\ket{\phi} = \bigotimes_{i=1}^n \ket{\phi_i} = \sum_{x=0}^{2^n-1} c_x \ket{x}$, where \(c_x \in \mathbb{C}\) and \(\sum |c_x|^2 = 1\). The index \(x \in \{0,1,...,2^{n}-1\}\) represents a possible observation of the quantum state \(\ket{\phi}\), which corresponds to the measurable output after a quantum measurement.

\subsection{\textcolor{black}{Quantum Gate}}
\label{quantum gates}
Quantum gates manipulate quantum states through unitary transformations, represented by matrices \(U\) satisfying \(U^\dagger U = I\). A quantum program consists of a sequence of gates applied to specific qubits. Quantum gates can be categorized on the basis of the number of qubits they operate on:
\begin{itemize}
    \item \textbf{Single-qubit:} Hadamard (H), Pauli (X, Y, Z), and others (S, Sdg, T, Tdg, U, P, Rx, Ry, Rz, Sx, Sxdg).
    \item \textbf{Two-qubit:} Controlled gates (CH, CS, CZ, CSdg, CP, CRx, CRy, CRz, CX) and others (SWAP, iSWAP).
    \item \textbf{Three-qubit:} Controlled gates (CCX, CCZ, CSWAP).
\end{itemize}


\textcolor{black}{Given a set of constraints, a fundamental challenge is to determine their satisfiability or to provide counterexamples.} 
This challenge serves as the motivation for implementing QSolver.


\section{\textcolor{black}{Approach}}
\label{sec:architeture}
This section describes the workflow and method of QSolver. We outline how QSolver represents quantum states, applies quantum gate transformations, defines constraints, and verifies results using assertions. The solver leverages dReal for constraint solving.

\subsection{\textcolor{black}{Workflow of QSolver}}
An overview of the QSolver architecture is presented in \autoref{fig:overview}. QSolver is implemented in Python and leverages the dReal SMT solver~\cite{dreal} for constraint solving. The key components of QSolver are described below:

\begin{figure*}[htb]
\centerline{\includegraphics[width=0.8\linewidth]{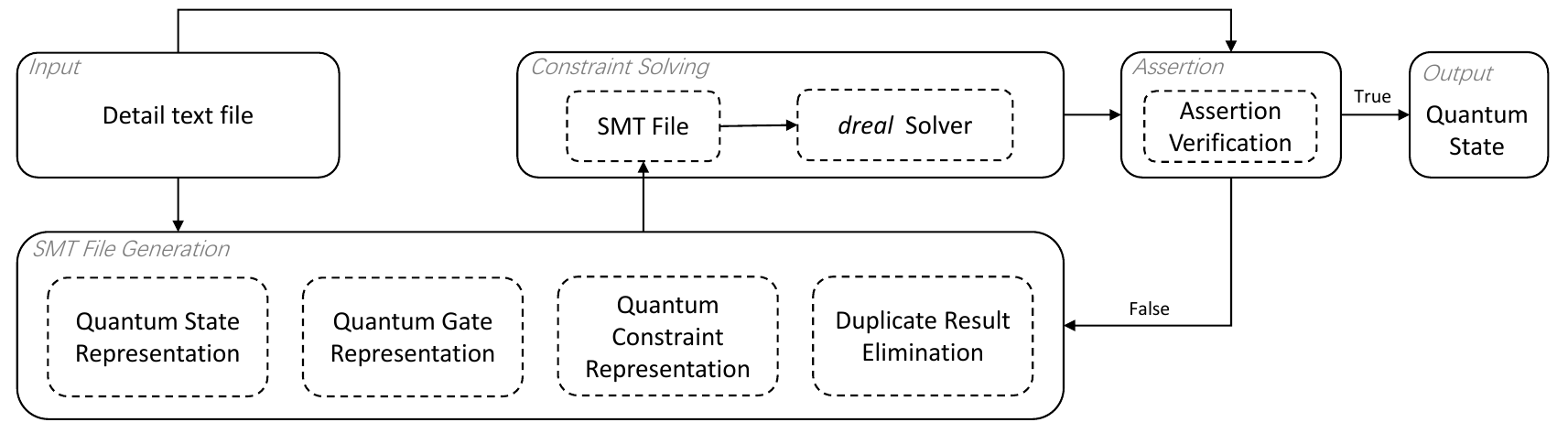}}
\vspace{-0.25cm}
\caption{Architecture of QSolver framework.}
\label{fig:overview}
\vspace{-3mm}
\end{figure*}

\begin{itemize}
\item \textbf{Input and Output:} To use QSolver, users must specify the number of qubits and provide a detailed text file. This file includes a list of quantum gates, a constraint flag indicating the type of constraint used, and the requirement to represent the expected output after measurement. For the requirement 1 in \autoref{fig:motivation}, the input text file specifying the required output states \texttt{010} and \texttt{101} of the qubits (index: [0,1,2]) is structured as follows:

\begin{lstlisting}
gates: [t(0) ; h(0) ; x(1) ; cx(1,2) ; cx(0,1)]
target_prob: [[0,1,2], [0,1,2], ['010','101']]
flag: "in"
\end{lstlisting}
The input for a multi-moment consists of a combination of multiple texts as described above.

\item \textbf{SMT File Generation:} QSolver consists of four SMT encoding modules responsible for handling different components:
    \begin{itemize}
        \item Quantum state representation (Section \ref{subsec:state-representation})
        \item Quantum gate transformations (Section \ref{subsec:gate-representation})
        \item Quantum constraints (Section \ref{subsec:constraint-represention})
        \item Avoidance of duplicate results (Section \ref{subsec:solving})
    \end{itemize}

\item \textbf{Constraint Solving:} Since the generated SMT file involves quadratic constraints, QSolver utilizes the dReal SMT solver to find solutions that satisfy the given constraints.

\item \textbf{Assertion Verification:} QSolver automatically generates an assertion program based on the requirements. The assertion program validates the solver's output by processing the results obtained from the dReal solver. If the validation fails, the computed result is passed to the Duplicate Result Elimination module to refine the solution.
\end{itemize}

The output of QSolver is a quantum state that satisfies the given constraints. For this example, the computed state is $[0j,\ 0j,\ 0j,\ 0j,\\(0.7071066697003517+0j),
\ -0.7071068926727259j,\ 0j,\ 0j]$

\subsection{\textcolor{black}{Quantum State Representation}}
\label{subsec:state-representation}
The solver represents a quantum system using classical variables, determined by the number of input qubits. For an $n$-qubit quantum system, we define the set $S = \{0,1,...,N\}$ where $N = 2^n - 1$, representing all possible observed values. The quantum state $\ket{\phi^{t}}$ at any computational step $t$ is expressed using real-valued classical variables $(a_i^t, b_i^t)$ as follows:
\begin{equation}
    \ket{\phi^{t}} \Rightarrow (a_0^{t}, a_1^{t}, \dots, a_N^{t}; b_0^{t}, b_1^{t}, \dots, b_N^{t}), i \in S.
\label{state_variables}
\end{equation}
Due to the inherent properties of quantum states, these classical variables must satisfy the normalization condition, ensuring the total probability sums to 1. Since quantum operations are typically unitary, it suffices to enforce normalization at the initial state, as follows:
\begin{equation}
    \textstyle \sum_{x\in S}((a_x^{0})^2 + (b_x^{0})^2) = 1.
\end{equation}
This constraint guarantees that subsequent quantum operations preserve the validity of the quantum state.

\subsection{\textcolor{black}{Quantum Gate Representation}}
\label{subsec:gate-representation}
QSolver applies quantum operations by updating the classical representation of quantum states. 
QSolver utilizes symbolic execution for each quantum operation to transform the representation of quantum states. To reduce the number of variables in the SMT file and accelerate the process, QSolver adopts a combinatorial approach that consolidates the quantum operations $(U_{1}, U_{2}...U_{t})$ preceding each constraint into a single symbolic transformation:


\vspace{-3mm}
\begin{equation}
    \ket{\phi^{0}}\overset{U_{all}}{\rightarrow} \ket{\phi^{t}}\Rightarrow (a_{i}^{0},b_{i}^{0}) \overset{U_{all}}{\rightarrow}(a_{i}^{t},b_{i}^{t}),\  U_{all}=U_{t}U_{t-1}...U_{1}.
\end{equation}


\subsection{\textcolor{black}{Quantum Constraint Representation}}
\label{subsec:constraint-represention}

\begin{table}[t]
\centering 
\caption{\textcolor{black}{Quantum constraints for five types of requirements and their corresponding SMT representations. Here, $S_r = \sum_{i \in O_r} i\otimes I_{B\setminus Q_m}$ represents the set of $n$-qubit states satisfying the observation condition, and $B=\{0,1,2,\dots,n-1\}$ denotes the set of qubit indices.}}
\label{constraint and representation}
\renewcommand{\arraystretch}{1.35} 
\setlength{\tabcolsep}{3pt} 
\begin{tabular}{|c|p{5.4cm}|}
\hline
\textbf{Flag} & \multicolumn{1}{c|}{\textbf{SMT Representation}} \\ \hline
$in$         & { $\textstyle \bigwedge_{x \in S\setminus S_r} (a_x^{final})^2+(b_x^{final})^2 = 0$}  \\ \hline
$not\_in$   & { $\textstyle \bigwedge_{x \in S_r} (a_x^{final})^2+(b_x^{final})^2 = 0$}  \\ \hline
$==$        & { $\textstyle \bigwedge_{x \in S} (|(a_x^{final})^2+(b_x^{final})^2 - D_x|\leq \delta)$}  \\ \hline
$>$         & { $\textstyle \bigwedge_{(x_s, p_s)\in P } ((a_{x_s}^{final})^2+(b_{x_s}^{final})^2 > p_s -\delta)$}  \\ \hline
$<$         & { $\textstyle \bigwedge_{(x_s, p_s)\in P} ((a_{x_s}^{final})^2+(b_{x_s}^{final})^2 \leq p_s +\delta)$}  \\ \hline
\end{tabular}
\vspace{-3mm}
\end{table}

After applying a sequence of quantum operations on a quantum state, it is necessary to characterize the probability distribution of that state through measurement. In our solver, we design five types of constraints, as summarized in \autoref{constraint and representation}, to describe the requirements for the probability distribution after measurement.

In the following formulas, we use $(a_{x}^{final}, b_{x}^{final})$ to denote the final quantum state before measurement. In actual implementation, $(a_{x}^{final}, b_{x}^{final})$ corresponds to $(a_{x}^{t}, b_{x}^{t})$ for the largest $t$ in \autoref{state_variables}.

\begin{itemize}

\item[$\bullet$] $in$/$not\_in$ (lines 1 and 2 in \autoref{constraint and representation}) for \textbf{In/Not\_In}: These constraints determine whether the measurement results of a specified set of qubit indices $Q_{m}$ belong to the expected observation set $O_{r}$. For example, in a 3-qubit program, measuring the first two qubits $Q_{m}=[0,1]$ should yield results in the set $O_r = \{'00', '11'\}$. The requirement is expressed as $[[0,1], [0,1], ['00', '11']]$.

\item[$\bullet$] $==$ (lines 3 in \autoref{constraint and representation}) for \textbf{Equal}: These constraints verify whether the output probability distribution $D$ satisfies the expected values within an allowable error margin $\delta$. For example, for a Bell state, the expected probability distribution $D$ is $[0.5, 0, 0, 0.5]$, which means $D_0=D_3=0.5$ and $D_1=D_2=0$.


\item[$\bullet$] $>$/$<$ (lines 4 and 5 in \autoref{constraint and representation}) for \textbf{Greater/Less}: These constraints check whether a particular observation $x_{s}$ of the quantum system exceeds or falls below a specified probability $p_{s}$. The requirement is represented as a list $P$ containing pairs $(x_{s}, p_{x})$. For example, $P=[[1, 0.3], [3, 0.4]]$ means that observations $x_s=1$ and $x_s=3$ should exceed or fall below the probabilities $p_s=0.3$ and $p_s=0.4$, respectively.

\end{itemize}

\subsection{\textcolor{black}{Solving by dReal}}
\label{subsec:solving}
After constructing the representations for quantum states, quantum operations, and quantum constraints, we establish the relational equations for $(a_{i}^{final},b_{i}^{final})$ in terms of $(a_{i}^{t}, b_{i}^{t})$. We then declare all used $(a_i^j, b_i^j)$ variables using \textit{declare-fun} and build a full SMT file, which is processed using the dReal SMT solver.

Due to inherent numerical errors in the dReal solver and the probabilistic nature of quantum program outputs, some results may not always satisfy the assertion conditions. To mitigate repeated incorrect results, we introduce a module to eliminate duplicate results. This module converts failing results into new constraints and appends them to the original SMT file, preventing the solver from generating the same incorrect outputs in subsequent iterations. Specifically, for any range $a_i^j \in [r_i^l, r_i^u]$ returned by the solver, we ensure that the next generated result differs by adding the constraint $(a_i^j<r_i^l-\varepsilon) \ \vee (a_i^j > r_i^u + \varepsilon)$. In the experiments, we set $\varepsilon=0.0005$.

\subsection{\textcolor{black}{Checking by Assertion}}
\label{subsec:assertion-checking}
\begin{table}[t]
\caption{\textcolor{black}{Assertion program generation for constraints. The assertion program performs multiple measurements and calculates the probability $mp_x$ of observing the expected outcome $x$. It then determines whether the assertion is satisfied by comparing $mp_x$ with the required value, allowing a tolerance of $\delta_{i}$. We set measurement times to 100000 and $\delta_{i}=0.05$.}}
\centering 
\label{assert} 

\begin{tabular}{|c|l|}
\hline
\ \ \textbf{Flag}\ \ \  & \multicolumn{1}{c|}{\textbf{Assertion}} \\ \hline
$in$         &    \   ${assert\ \  \textstyle (\sum_{x \in S_{r} } mp_{x}) > 0.95-\delta_{i}} $      \\ \hline
$not\_in$    &   \   $assert\ \  \textstyle (\sum_{x \in S_{r} } mp_{x}) < 0.05+\delta_{i}$       \\ \hline
$==$     &   \    $assert\ \  \textstyle \bigwedge_{x \in S} (|mp_{x} - D_x|\le \delta_{i})$      \\ \hline
$>$     &  \    $assert \ \  \textstyle \bigwedge_{s \in P \wedge  s=(x_s, p_s)} (mp_{x_{s}} > p_s -\delta_{i})$  \\ \hline
$<$       &  \    $assert \ \  \textstyle \bigwedge_{s \in P \wedge  s=(x_s, p_s)} (mp_{x_{s}} < p_s -\delta_{i})$  \\ \hline
\end{tabular}
\vspace{-3mm}
\end{table}

To verify that dReal-generated results satisfy the expected constraints, we automate the generation of assertion programs, as summarized in \autoref{assert}. Based on user-provided input information, QSolver generates the corresponding Qiskit program~\cite{qiskit} to validate the dReal output. For each constraint, QSolver retains only the preceding quantum operations and utilizes assertion statements for verification.






\begin{table*}[h]
\scriptsize{
\caption{\textcolor{black}{Experimental Results of Different 29 Quantum Gates. For each quantum gate, we perform five constraint types with qubit numbers 1, 2, 3, and 4, respectively. \textbf{$Time(i)$}: The average generation time per result is $Time$ seconds and the $i_{th}$ result generated by QSolver satisfies the assertion. \textbf{NF}: QSolver generated 10 results that failed to satisfy the assertion. \textbf{TO}: Calculation time exceeds 1000s. \textbf{NE}: Not exist.}}
\label{different_gates}
\begin{tabular}{cc|cccccccccccccccccccc}
\hline
\multicolumn{2}{c|}{\multirow{3}{*}{\begin{tabular}[c]{@{}c@{}}\textbf{Quantum}\\ \textbf{Gates}\end{tabular}}}            & \multicolumn{20}{c}{\textbf{Constraint Types}}                                                                                                                                                                                                                                                                                                                                                                                                                                                      \\ \cline{3-22} 
\multicolumn{2}{c|}{}                                                                                    & \multicolumn{4}{c|}{$in$}                                                          & \multicolumn{4}{c|}{$not\_in$}                                                     & \multicolumn{4}{c|}{$==$}                                                                                          & \multicolumn{4}{c|}{$>$}                                                      & \multicolumn{4}{c}{$<$}                                    \\ \cline{3-22} 
\multicolumn{2}{c|}{}                           & 1                                         & 2   & 3   & \multicolumn{1}{c|}{4}   & 1                                         & 2   & 3    & \multicolumn{1}{c|}{4}   & 1                                         & 2   & 3   & \multicolumn{1}{c|}{4}  & 1                                         & 2   & 3   & \multicolumn{1}{c|}{4}   & 1                                         & 2   & 3   & 4   \\ \hline
\multicolumn{1}{c|}{\multirow{15}{*}{\begin{tabular}[c]{@{}c@{}}1\\ qubit\\ gates\end{tabular}}} & H     & 0.01(1)                                       & 0.02(1) & 557(1) & \multicolumn{1}{c|}{TO}  & 0.01(1)                                       & 0.03(1) & 0.26(4) & \multicolumn{1}{c|}{NF}  & 0.02(1)                                       & 0.03(1) & 0.08(1)  & \multicolumn{1}{c|}{0.44(1)} & 0.02(1)                                       & 56.7(1) & 0.13(1) & \multicolumn{1}{c|}{TO}  & 0.02(1)                                       & 0.04(1) & 6.61(1) & TO  \\
\multicolumn{1}{c|}{}                                                                            & X     & 0.01(1)                                       & 0.01(1) & 0.03(1) & \multicolumn{1}{c|}{0.06(4)} & 0.01(1)                                       & 0.02(3) & 0.02(1) & \multicolumn{1}{c|}{0.07(1)} & 0.01(1)                                       & 0.02(1) & 0.03(1)  & \multicolumn{1}{c|}{0.10(1)} & 0.01(1)                                       & 0.02(1) & 0.03(1) & \multicolumn{1}{c|}{TO}  & 0.01(1)                                       & 0.01(1) & 0.02(1) & 0.06(1) \\
\multicolumn{1}{c|}{}                                                                            & S     & 0.01(1)                                       & NF  & 0.03(3) & \multicolumn{1}{c|}{0.07(1)} & 0.01(1)                                       & NF  & 0.04(4) & \multicolumn{1}{c|}{0.07(1)} & 0.01(1)                                       & 0.02(1) & 0.03(1)  & \multicolumn{1}{c|}{0.11(1)} & 0.01(1)                                       & 0.02(1) & 0.03(1) & \multicolumn{1}{c|}{0.15(1)} & 0.01(1)                                       & 0.01(1) & 0.03(1) & 0.09(1) \\
\multicolumn{1}{c|}{}                                                                            & Z     & 0.01(1)                                       & 0.02(3) & 0.03(3) & \multicolumn{1}{c|}{NF}  & 0.01(1)                                       & 0.02(3) & 0.02(1) & \multicolumn{1}{c|}{0.10(3)} & 0.01(1)                                       & 0.02(1) & 0.03(1)  & \multicolumn{1}{c|}{0.10(1)} & 0.02(1)                                       & 0.02(1) & 112(1) & \multicolumn{1}{c|}{TO}  & 0.02(1)                                       & 0.02(1) & 0.02(1) & 0.07(1) \\
\multicolumn{1}{c|}{}                                                                            & Y     & 0.01(1)                                       & 0.01(1) & 0.02(1) & \multicolumn{1}{c|}{0.07(1)} & 0.01(1)                                       & 0.02(3) & 0.03(1) & \multicolumn{1}{c|}{0.07(1)} & 0.01(1)                                       & 0.02(1) & 0.03(1)  & \multicolumn{1}{c|}{0.10(1)} & 0.01(1)                                       & 0.01(1) & 0.11(1) & \multicolumn{1}{c|}{0.39(1)} & 0.01(1)                                       & 0.02(1) & 0.02(1) & 0.08(1) \\
\multicolumn{1}{c|}{}                                                                            & Sdg   & 0.01(1)                                       & 0.02(3) & 0.03(3) & \multicolumn{1}{c|}{0.06(1)} & 0.01(1)                                       & 0.02(1) & 0.02(1) & \multicolumn{1}{c|}{0.06(1)} & 0.01(1)                                       & 0.02(1) & 0.03(1)  & \multicolumn{1}{c|}{0.09(1)} & 0.01(1)                                       & 0.02(1) & 0.35(1) & \multicolumn{1}{c|}{8.74(1)} & 0.01(1)                                       & 0.02(1) & 0.02(1) & 0.07(1) \\
\multicolumn{1}{c|}{}                                                                            & T     & 0.01(1)                                       & 0.02(3) & 0.04(4) & \multicolumn{1}{c|}{0.06(1)} & 0.01(1)                                       & 0.01(1) & 0.02(1) & \multicolumn{1}{c|}{0.09(1)} & 0.01(1)                                       & 0.02(1) & 0.03(1)  & \multicolumn{1}{c|}{0.11(1)} & 0.01(1)                                       & 0.02(1) & 0.20(1) & \multicolumn{1}{c|}{8.59(1)} & 0.01(1)                                       & 0.02(1) & 0.02(1) & 0.10(1) \\
\multicolumn{1}{c|}{}                                                                            & Tdg   & 0.01(1)                                       & 0.03(4) & NF  & \multicolumn{1}{c|}{0.11(2)} & 0.01(1)                                       & 0.44(4) & 0.04(1) & \multicolumn{1}{c|}{0.49(1)} & 0.01(1)                                       & 0.02(1) & 0.03(1)  & \multicolumn{1}{c|}{0.12(1)} & 0.01(1)                                       & 0.26(1) & TO  & \multicolumn{1}{c|}{TO}  & 0.01(1)                                       & 0.02(1) & 0.03(1) & 1.75(1) \\
\multicolumn{1}{c|}{}                                                                            & U     & 0.01(1)                                       & 0.03(4) & 2.19(2) & \multicolumn{1}{c|}{TO}  & 0.02(1)                                       & 0.04(1) & 1.56(1) & \multicolumn{1}{c|}{TO}  & 0.01(1)                                       & 0.02(1) & 932(1)  & \multicolumn{1}{c|}{0.26(1)} & 0.02(1)                                       & 0.04(1) & TO  & \multicolumn{1}{c|}{TO}  & 0.02(1)                                       & 0.03(1) & TO  & TO  \\
\multicolumn{1}{c|}{}                                                                            & P     & 0.02(1)                                       & 0.01(1) & NF  & \multicolumn{1}{c|}{0.06(1)} & 0.01(1)                                       & 0.02(1) & 0.04(1) & \multicolumn{1}{c|}{0.08(1)} & 0.01(1)                                       & 0.02(1) & 0.04(1)  & \multicolumn{1}{c|}{0.09(1)} & 0.01(1)                                       & 0.02(1) & 0.27(1) & \multicolumn{1}{c|}{TO}  & 0.01(1)                                       & 0.02(1) & 0.06(1) & 0.07(1) \\
\multicolumn{1}{c|}{}                                                                            & Rx    & 0.01(1)                                       & 0.02(1) & 0.04(1) & \multicolumn{1}{c|}{0.37(1)} & 0.01(1)                                       & 0.05(1) & 0.05(2) & \multicolumn{1}{c|}{TO}  & 0.01(1)                                       & 0.02(1) & 0.04(1)  & \multicolumn{1}{c|}{0.17(1)} & 0.01(1)                                       & 0.02(1) & 265(1) & \multicolumn{1}{c|}{TO}  & 0.01(1)                                       & 0.03(1) & 0.05(1) & TO  \\
\multicolumn{1}{c|}{}                                                                            & Ry    & 0.02(1)                                       & 0.02(1) & 0.04(1) & \multicolumn{1}{c|}{0.13(1)} & 0.01(1)                                       & 0.05(1) & 0.04(1) & \multicolumn{1}{c|}{0.17(1)} & 0.02(1)                                       & 0.03(1) & 0.05(1)  & \multicolumn{1}{c|}{0.17(1)} & 0.01(1)                                       & 0.06(1) & TO  & \multicolumn{1}{c|}{TO}  & 0.02(1)                                       & 0.04(1) & 0.48(1) & TO  \\
\multicolumn{1}{c|}{}                                                                            & Rz    & 0.01(1)                                       & 0.02(3) & 0.03(1) & \multicolumn{1}{c|}{TO}  & 0.01(1)                                       & 0.03(8) & 3.47(1) & \multicolumn{1}{c|}{TO}  & 0.01(1)                                       & 0.02(1) & 0.04(1)  & \multicolumn{1}{c|}{0.17(1)} & 0.01(1)                                       & 6.85(1) & 109(1) & \multicolumn{1}{c|}{TO}  & 0.01(1)                                       & 0.02(1) & 24.1(1) & 583(1) \\
\multicolumn{1}{c|}{}                                                                            & Sx    & 0.01(1)                                       & 2.28(4) & 3.22(1) & \multicolumn{1}{c|}{TO}  & 0.01(1)                                       & 416(2) & 25.8(1) & \multicolumn{1}{c|}{TO}  & 0.02(1)                                       & 0.03(1) & TO   & \multicolumn{1}{c|}{TO} & 0.02(1)                                       & 2.71(1) & TO  & \multicolumn{1}{c|}{TO}  & 0.02(1)                                       & TO  & TO  & TO  \\
\multicolumn{1}{c|}{}                                                                            & Sxdg  & 0.02(1)                                       & 0.03(1) & 44.3(1) & \multicolumn{1}{c|}{TO}  & 0.01(1)                                       & 0.23(2) & TO  & \multicolumn{1}{c|}{TO}  & 0.02(1)                                       & 0.03(1) & 425(1)  & \multicolumn{1}{c|}{TO} & 0.02(1)                                       & 0.27(1) & TO  & \multicolumn{1}{c|}{TO}  & 0.02(1)                                       & 0.44(1) & TO  & TO  \\ \hline
\multicolumn{1}{c|}{\multirow{11}{*}{\begin{tabular}[c]{@{}c@{}}2\\ qubit\\ gates\end{tabular}}} & CH    & \multicolumn{1}{c|}{\multirow{11}{*}{NE}} & 0.03(1) & 0.07(1) & \multicolumn{1}{c|}{0.21(1)} & \multicolumn{1}{c|}{\multirow{11}{*}{NE}} & 0.02(1) & 0.13(2) & \multicolumn{1}{c|}{TO}  & \multicolumn{1}{c|}{\multirow{11}{*}{NE}} & 0.03(1) & 0.09(10) & \multicolumn{1}{c|}{NF}  & \multicolumn{1}{c|}{\multirow{11}{*}{NE}} & 0.03(1) & TO  & \multicolumn{1}{c|}{TO}  & \multicolumn{1}{c|}{\multirow{11}{*}{NE}} & 0.08(2) & 2.95(1) & TO  \\
\multicolumn{1}{c|}{}                                                                            & CS    & \multicolumn{1}{c|}{}                     & 0.01(1) & 0.02(1) & \multicolumn{1}{c|}{0.07(5)} & \multicolumn{1}{c|}{}                     & 0.02(3) & 0.03(1) & \multicolumn{1}{c|}{0.06(1)} & \multicolumn{1}{c|}{}                     & 0.02(1) & 0.03(1)  & \multicolumn{1}{c|}{0.11(1)} & \multicolumn{1}{c|}{}                     & 0.02(1) & 52.2(1) & \multicolumn{1}{c|}{882(1)} & \multicolumn{1}{c|}{}                     & 0.02(1) & 0.03(1) & 0.06(1) \\
\multicolumn{1}{c|}{}                                                                            & CZ    & \multicolumn{1}{c|}{}                     & 0.02(3) & 0.02(1) & \multicolumn{1}{c|}{0.07(5)} & \multicolumn{1}{c|}{}                     & 0.02(3) & 0.03(4) & \multicolumn{1}{c|}{0.06(1)} & \multicolumn{1}{c|}{}                     & 0.02(1) & 0.04(1)  & \multicolumn{1}{c|}{0.11(1)} & \multicolumn{1}{c|}{}                     & 0.02(1) & 384(1) & \multicolumn{1}{c|}{TO} & \multicolumn{1}{c|}{}                     & 0.02(1) & 0.02(1) & 0.07(1) \\
\multicolumn{1}{c|}{}                                                                            & CSdg  & \multicolumn{1}{c|}{}                     & 0.02(3) & 0.02(1) & \multicolumn{1}{c|}{0.08(3)} & \multicolumn{1}{c|}{}                     & NF  & 0.02(1) & \multicolumn{1}{c|}{0.06(1)} & \multicolumn{1}{c|}{}                     & 0.02(1) & 0.03(1)  & \multicolumn{1}{c|}{0.09(1)} & \multicolumn{1}{c|}{}                     & 0.01(1) & 798(1) & \multicolumn{1}{c|}{1.06(1)} & \multicolumn{1}{c|}{}                     & 0.01(1) & 0.02(1) & 0.07(1) \\
\multicolumn{1}{c|}{}                                                                            & CP    & \multicolumn{1}{c|}{}                     & 0.01(1) & 0.03(1) & \multicolumn{1}{c|}{0.08(3)} & \multicolumn{1}{c|}{}                     & 0.01(1) & 0.04(5) & \multicolumn{1}{c|}{0.06(1)} & \multicolumn{1}{c|}{}                     & 0.02(1) & 0.03(1)  & \multicolumn{1}{c|}{0.09(1)} & \multicolumn{1}{c|}{}                     & 0.02(1) & 0.06(1) & \multicolumn{1}{c|}{0.13(1)} & \multicolumn{1}{c|}{}                     & 0.02(1) & 0.03(1) & 0.07(1) \\
\multicolumn{1}{c|}{}                                                                            & CRx   & \multicolumn{1}{c|}{}                     & 0.02(2) & 0.05(1) & \multicolumn{1}{c|}{0.14(1)} & \multicolumn{1}{c|}{}                     & 0.03(2) & 0.06(2) & \multicolumn{1}{c|}{0.14(1)} & \multicolumn{1}{c|}{}                     & 0.03(8) & 0.06(4)  & \multicolumn{1}{c|}{0.17(1)} & \multicolumn{1}{c|}{}                     & 0.08(1) & TO  & \multicolumn{1}{c|}{TO}  & \multicolumn{1}{c|}{}                     & 0.02(1) & 0.07(1) & 0.14(1) \\
\multicolumn{1}{c|}{}                                                                            & CRy   & \multicolumn{1}{c|}{}                     & 0.02(1) & 0.04(1) & \multicolumn{1}{c|}{0.15(2)} & \multicolumn{1}{c|}{}                     & 0.03(2) & 0.06(1) & \multicolumn{1}{c|}{0.17(1)} & \multicolumn{1}{c|}{}                     & 0.03(2) & 0.07(6)  & \multicolumn{1}{c|}{0.17(1)} & \multicolumn{1}{c|}{}                     & 0.02(1) & TO  & \multicolumn{1}{c|}{TO}  & \multicolumn{1}{c|}{}                     & 0.03(1) & 0.25(1) & 0.15(1) \\
\multicolumn{1}{c|}{}                                                                            & CRz   & \multicolumn{1}{c|}{}                     & 0.03(7) & 0.04(1) & \multicolumn{1}{c|}{0.16(1)} & \multicolumn{1}{c|}{}                     & 0.03(8) & 0.15(2) & \multicolumn{1}{c|}{NF}  & \multicolumn{1}{c|}{}                     & 0.02(1) & 0.05(1)  & \multicolumn{1}{c|}{0.17(1)} & \multicolumn{1}{c|}{}                     & 0.03(1) & 904(1) & \multicolumn{1}{c|}{TO}  & \multicolumn{1}{c|}{}                     & 0.03(1) & 0.04(1) & 3.64(1) \\
\multicolumn{1}{c|}{}                                                                            & SWAP  & \multicolumn{1}{c|}{}                     & 0.02(1) & 0.03(1) & \multicolumn{1}{c|}{0.08(5)} & \multicolumn{1}{c|}{}                     & 0.02(3) & 0.02(1) & \multicolumn{1}{c|}{0.06(1)} & \multicolumn{1}{c|}{}                     & 0.02(1) & 0.03(1)  & \multicolumn{1}{c|}{0.08(1)} & \multicolumn{1}{c|}{}                     & 0.02(1) & 0.05(1) & \multicolumn{1}{c|}{TO}  & \multicolumn{1}{c|}{}                     & 0.01(1) & 0.02(1) & 0.06(1) \\
\multicolumn{1}{c|}{}                                                                            & iSWAP & \multicolumn{1}{c|}{}                     & 0.01(1) & 0.03(1) & \multicolumn{1}{c|}{0.06(3)} & \multicolumn{1}{c|}{}                     & 0.02(3) & 0.04(4) & \multicolumn{1}{c|}{0.06(1)} & \multicolumn{1}{c|}{}                     & 0.02(1) & 0.03(1)  & \multicolumn{1}{c|}{0.09(1)} & \multicolumn{1}{c|}{}                     & 0.02(1) & 0.05(1) & \multicolumn{1}{c|}{0.23(1)} & \multicolumn{1}{c|}{}                     & 0.01(1) & 0.02(1) & 0.09(1) \\
\multicolumn{1}{c|}{}                                                                            & CX    & \multicolumn{1}{c|}{}                     & 0.01(1) & 0.02(1) & \multicolumn{1}{c|}{0.06(1)} & \multicolumn{1}{c|}{}                     & 0.02(1) & 0.02(1) & \multicolumn{1}{c|}{0.06(1)} & \multicolumn{1}{c|}{}                     & 0.02(1) & 0.03(1)  & \multicolumn{1}{c|}{0.12(1)} & \multicolumn{1}{c|}{}                     & 0.02(1) & 0.12(1) & \multicolumn{1}{c|}{TO}  & \multicolumn{1}{c|}{}                     & 0.02(1) & 0.02(1) & 0.08(1) \\ \hline
\multicolumn{1}{c|}{\multirow{3}{*}{\begin{tabular}[c]{@{}c@{}}3\\ qubit\\ gates\end{tabular}}}  & CCX   & \multicolumn{2}{c|}{\multirow{3}{*}{NE}}        & 0.02(1) & \multicolumn{1}{c|}{0.07(1)} & \multicolumn{2}{c|}{\multirow{3}{*}{NE}}        & 0.03(3) & \multicolumn{1}{c|}{0.05(1)} & \multicolumn{2}{c|}{\multirow{3}{*}{NE}}        & 0.03(1)  & \multicolumn{1}{c|}{0.09(1)} & \multicolumn{2}{c|}{\multirow{3}{*}{NE}}        & 1.44(1) & \multicolumn{1}{c|}{TO}  & \multicolumn{2}{c|}{\multirow{3}{*}{NE}}        & 0.03(1) & 0.07(3) \\
\multicolumn{1}{c|}{}                                                                            & CCZ  & \multicolumn{2}{c|}{}                           & 0.04(3) & \multicolumn{1}{c|}{0.06(1)} & \multicolumn{2}{c|}{}                           & 0.04(3) & \multicolumn{1}{c|}{0.06(1)} & \multicolumn{2}{c|}{}                           & 0.03(1)  & \multicolumn{1}{c|}{0.12(1)} & \multicolumn{2}{c|}{}                           & 0.03(1) & \multicolumn{1}{c|}{1.33(1)} & \multicolumn{2}{c|}{}                           & 0.02(1) & 0.06(1) \\
\multicolumn{1}{c|}{}                                                                            & CSWAP & \multicolumn{2}{c|}{}                           & 0.03(3) & \multicolumn{1}{c|}{0.06(1)} & \multicolumn{2}{c|}{}                           & 0.02(1) & \multicolumn{1}{c|}{0.07(4)} & \multicolumn{2}{c|}{}                           & 0.03(1)  & \multicolumn{1}{c|}{0.08(1)} & \multicolumn{2}{c|}{}                           & 0.04(1) & \multicolumn{1}{c|}{214(1)} & \multicolumn{2}{c|}{}                           & 0.02(1) & 0.06(1) \\
\hline
\end{tabular}
}
\vspace{-3mm}
\end{table*}

\section{Experiments}
\label{sec:experiments}
\textcolor{black}{The performance of QSolver\footnote{\url{https://github.com/Xzore19/QSolver}} was evaluated through two experiments. All experiments were carried out on a computer running Ubuntu 22.04 LTS, equipped with an Intel Xeon E5-2620v4 processor, 32GB of memory, and a 2TB HDD.}

\vspace{-0.5mm}
\subsection{Performance on Different Quantum Gates} 
\textcolor{black}{We conducted experiments on the most commonly used quantum gates, and the results are presented in \autoref{different_gates}.}
\textcolor{black}{Experiment results indicate that QSolver can effectively process 29 widely used quantum gates in Section~\ref{quantum gates} and generate quantum states that satisfy the specified conditions.} 
\textcolor{black}{The assertion programs we generated can effectively provide the testing environment. Moreover, since the constraints are randomly generated, some quantum operations may produce SMT files that pose significant solving challenges for dReal. We hypothesize that this could be due to either inherent instabilities in dReal or the extremely small solution space.}

\subsection{Performance on Different Gate Sizes} 
We evaluated QSolver's performance across various quantum circuit sizes, and the results are presented in \autoref{different_sizes}. 
Our findings indicate that QSolver can effectively handle \textcolor{black}{quantum programs with varying circuit sizes}.
\textcolor{black}{Furthermore, in the multi-moment scenario, we deliberately constructed some conflicting constraints, and QSolver can effectively return \textit{unsat} as an indication.}

\begin{table}[htb]
\centering
\caption{\textcolor{black}{Experimental Results for Different Quantum Gate Sizes. The experiment was conducted on quantum programs with 5 and 10 random gates, and qubit numbers 1, 2, and 3, respectively.} $(i)$ is the number of inputs for each program.}
\label{different_sizes}
\resizebox{\linewidth}{!}{
\begin{tabular}{c|ccccccccccccccc}
\hline
\multirow{3}{*}{\textbf{Gates}} & \multicolumn{15}{c}{\textbf{Flag}}                                                                                                                                                          \\ \cline{2-16} 
                                & \multicolumn{3}{c|}{$in$}              & \multicolumn{3}{c|}{$not\_in$}         & \multicolumn{3}{c|}{$==$}              & \multicolumn{3}{c|}{$>$}  & \multicolumn{3}{c}{$<$} \\ \cline{2-16} 
                                & 1   & 2   & \multicolumn{1}{c|}{3}   & 1   & 2   & \multicolumn{1}{c|}{3}   & 1   & 2   & \multicolumn{1}{c|}{3}   & 1   & 2   & \multicolumn{1}{c|}{3}   & 1         & 2        & 3        \\ \hline
5                               & (1) & (1) & \multicolumn{1}{c|}{(1)} & (1) & (1) & \multicolumn{1}{c|}{(1)} & (1) & (1) & \multicolumn{1}{c|}{(1)} & (1) & (1) & \multicolumn{1}{c|}{(1)} & (1)       & (2)      & (1)      \\
10                              & (1) & (1) & \multicolumn{1}{c|}{(1)} & (1) & (2) & \multicolumn{1}{c|}{(2)} & (1) & (2) & \multicolumn{1}{c|}{(1)} & (1) & (1) & \multicolumn{1}{c|}{(1)} & (1)       & (1)      & (1)     
\end{tabular}
}
\vspace{-5mm}
\end{table}

\section{\textcolor{black}{Future Plans}}
In the future, we aim to enhance this framework through further refinements and the incorporation of additional functionalities.

\textit{Efficiency.} \textcolor{black}{Theoretically, QSolver is applicable to an arbitrary number of qubits. However, limited by the efficiency of constraint solving, our experiments were limited to 1–3 qubits.}
We aim to address the challenge of long solver runtimes by exploring optimization techniques, such as leveraging the stabilizer.

\textit{Extensibility.} We plan to extend QSolver to incorporate additional quantum constraints, such as those for analyzing quantum entanglement and superposition properties, thereby enhancing our understanding of the behavior of quantum programs. 
\textcolor{black}{We also plan to extend the assertion program to other quantum languages, such as Cirq~\cite{cirq} and Q\#~\cite{QsSpec2020}.}

\section{Conclusion}
\label{sec:conclusion}
\textcolor{black}{In this paper, we presented QSolver, a constraint solver for quantum programs that enables the effective specification and verification of quantum state constraints. QSolver provides a structured framework for handling five types of quantum constraints and includes an automated assertion generation module to validate the generated states. Experimental results demonstrate that QSolver efficiently supports 29 commonly used quantum gates and shows good scalability across quantum programs of different sizes.}

\begin{acks}
This work was supported in part by JST SPRING Grant No.\ JPMJSP2136, and JSPS KAKENHI Grants No.\ JP23K28062, No.\ JP24K\allowbreak14908, and No.\ JP24K02920.
\end{acks}

\bibliographystyle{ACM-Reference-Format}
\bibliography{ref}

\end{document}